\documentstyle[12pt]{article}
\def\CS{{\cal S}}
\def\CA{{\cal A}}
\def\SP{{\sf P}}
\def\CP{{\cal P}}
\def\BCP{\bar{{\cal P}}}
\def\CH{{\cal H}}

\begin{document}
\begin{flushright}   FIAN/TD/95-13 \\
                hep-ph/9506203\\
                June 1995 \vspace{3ex}
\end{flushright}

\vspace*{1cm}

\begin{center}
{\Large ON THE PERTURBATIVE EQUIVALENCE BETWEEN }
\end{center}

\vspace{0.41cm}

\begin{center}
{\Large THE HAMILTONIAN AND LAGRANGIAN QUANTIZATIONS}
\end{center}

\vspace{1cm}
\begin{center}
I. A. Batalin and I. V. Tyutin
\end{center}

\begin{center}
{\it Lebedev Physical Institute, 117924 Moscow Russia}
\end{center}

\vspace{6cm}
\begin{center}
{\large Abstract}
\end{center}

The Hamiltonian (BFV) and Lagrangian (BV) quantization schemes are proved to be
equivalent perturbatively to each other. It is shown in particular that the
quantum master equation being treated perturbatively possesses a local formal
solution.

\newpage

\section{Introduction}

The   Hamiltonian (BFV) [l-4] and Lagrangian (BV) [5 -- 7] quantization schemes
are the most popular ones presently. Both these approaches are shown to be
equivalent to each other as applied to all the known field--theoretic  examples
(Yang-Mills theory, gravity, etc.). However it would be very interesting in
principle to compare these approaches in the most general case.

It is  relevant to  elucidate here what we mean by the term "equivalence". As
it is well--known [8], the general solution to generating equations of the
BFV--method is determined up to a canonical transformation and a choice of
gauge--fixing fermion, so that all such solutions are physically-equivalent. On
the other hand, in the BV--method the general solution to the classical master
equation is determined [10, 6] up to an anticanonical transformation, and all
such solutions are also physically--equivalent. The general solution to the
quantum master equation is determined [11, 12] up to an anticanonical
transformation accompanied by adding gauge--invariant "quantum corrections"
$\sim O(\hbar)$ to the initial classical action. These "quantum corrections"
result in the fact that different solutions to the quantum master equation may
appear to be physically--nonequivalent. Therefore, when reasoning on the
physical equivalence between two different quantum  master actions (or even
between two different quantum theories at all) we mean the following: there
exists a solution to the first quantum master equation (or an action of the
first theory), which is physically--equivalent to the given solution to the
second master equation (or to the given action of the second theory).

In principle, it has been shown in Ref. [13] that there exists an effective
action depending only on the subset of phase variables of the BFV--quantization
(in particular, on the variables of the Lagrangian BV--quantization only), as
well  as on the corresponding antivariables, which effective action satisfies
the quantum master equation. It remains, however, unclear what is the relation
between the mentioned effective action and the initial one. The equivalence
between the two approaches has been proved in Ref. [14, 15] for theories with
first--class constraints only (and only up to a local measure as for Ref.
[15]).
As a consequence of the result, it was shown that the quantum master equation
as
applied to such theories possesses a local solution. In Ref. [16] the general
case of theories with constraints of the both classes has been considered.
However, the equivalence of the two approaches to the quantization problem was
proved here up to a local measure. In the present paper we eliminate all the
above mentioned restrictions and prove the formal perturbative equivalence
between the Hamiltonian and Lagrangian quantizations as applied to an arbitrary
classical theory. As a basic point, we generalize the result of Ref. [17] about
the quantum equivalence of the classically--equivalent theories.

We treat of the classical equivalence in the following sense. Let the motion
equations

$$
{\delta\CS(x,y)\over\delta y}=0
$$
determine $y$ uniquely to be some functions $f$ of $x$: $y=f(x)$. Then the
actions $\CS(x,y)$ and $\CS_1(x)\equiv\CS(x,f(x))$ are  classically-equivalent.
Usually we also apply the same terminology directly to the
theories with actions $\CS$ and $\CS_1$.

It was shown in Ref. [17] that, given a solution to the master equation for
the theory with classical action $\CS_1(x)$, then there exists a
physically--equivalent solution to the master equation for the theory with
classical action $\CS(x,y)$.

In Section 2 we extend the result of Ref. [17] to cover the case of quantum
master equations and, what is more, we prove the coresponding vice--versa
result
as well.

In Section 3, given the classical action $\CS_L$, we construct the
classically--equivalent action $\CS_{1H}$ to be convenient to us, which has the
form of a Hamiltonian action with first--class constraints only.

In Section 4, by making use of the method of Ref.[14], we prove the existence
of a solution to the master equation for the theory with classical action
$\CS_{1H}$, which solution is in fact equivalent to the Dirac quantization of
the classical theory $\CS_L$. It follows then from the results of Section 2
that there exists a solution to the quantum master equation for the theory with
classical actions $\CS_L$, which solution is equivalent to the Dirac
quantization and, thereby, is equivalent to the BFV--one.

For the sake of simplicity, in what follows below we restrict ourselves by the
case of purely bosonic initial Lagrangian variables.

\section{Quantum equivalence of classically--equivalent theories in Lagrangian
approach}

Let us consider the action $\CS(\varphi)\equiv\CS(x,y)$ such that the equations
of
motion

$$
{\delta\CS\over\delta y^a}=0
\eqno{(2.1)}$$
determine $y^a$ uniquely to be some functions $f^a$ of $x$:

$$
{\delta\CS\over\delta y^a}=0\quad \leftrightarrow \quad
y^a=f^a(x).  \eqno{(2.2)}$$ The reduced action $\CS_1(x)$:

$$
\CS_1(x)\equiv\CS(x,f(x))
\eqno{(2.3)}$$
is physically--equivalent to the initial one $\CS(x,y)$.

If one supposes the initial action $\CS(x,y)$ to be gauge--invariant, with

$$
R^A_\alpha(x,y),\quad\varphi^A\equiv(x^i,y^a),
\eqno{(2.4)}$$
being the corresponding gauge generators, then the reduced functions

$$
R^i_{1\alpha}(x)\equiv R^i_\alpha(x,f(x))
\eqno{(2.5)}$$
serve as gauge generators to the reduced action $\CS_1(x)$, and vice versa as
well.

Let us derive a property of the action $\CS(x,y)$ to be important in what
follows
below. By making the following unimodular change of variables

$$
\varphi^A\quad \rightarrow \quad \varphi^{\prime A}=(x^i,z^a=y^a-f^a(x))
\eqno{(2.6)}$$
let us transform the action to be represented in terms of new variables as

$$
\CS(x,y)=\CS^\prime(x,z)=\CS_1(x)+{1\over2}\int dtz^a\Lambda_{ab}(x,z)z^b
\eqno{(2.7)}$$
with $\Lambda_{ab}(x,z)$ being finite--order differential operator (FODO) with
coefficients depending on $x$, $z$.

Lemma 1. There exists a change of variables

$$
z^a \quad \rightarrow \quad Y^a=\mu^a_b(x,z)z^b, \quad
\mu^a_b(x,z)=\delta^a_b+O(\varphi^\prime),
 \eqno{(2.8)}$$
with $\mu^a_b(x,z)$ being FODO to each finite order in $x$,
$z$, such that the action takes the form

$$
\CS^\prime(x,z)=\CS(x)+{1\over2}\int dtY^a\Lambda^{(0)}_{ab}Y^b
\eqno{(2.9)}$$
where

$$
\Lambda^{(0)}_{ab}=\Lambda_{ab}(0,0).
\eqno{(2.10)}$$

Proof. Let us suppose that we have constructed FODO $\mu^{(n)}{}^a_b(x,z)$ such
that

$$
\CS^\prime(x,z)=\CS_1(x)+{1\over2}\int dt[Y^{(n)a}\Lambda^{(0)}_{ab}Y^{(n)b}+
z^a\delta_{n+1}\Lambda_{ab}(x,z)z^b]
\eqno{(2.11)}$$
where

$$
Y^{(n)a}=\mu^{(n)}{}^a_b(x,z)z^b
\eqno{(2.12)}$$

$$
\delta_{n+1}\Lambda_{ab}(x,z)=O((\varphi^\prime)^{n+1})
\eqno{(2.13)}$$
with $\delta_{n+1}\Lambda_{ab}$ being FODO. Let us choose

$$
Y^{(n+1)a}=\Bigl(\mu^{(n)}{}^a_b(x,z)+{1\over2}\overrightarrow{\Sigma}{}^{ac}
\delta_{n+1}\overrightarrow{\Lambda}_{cb}(x,z)\Bigr)z^b,
\eqno{(2.14)}$$
where, by definition, we set

$$
\int dt\varphi^aQ_{ab}\psi^b=\int dt\varphi^a\psi_{Qa}=
\int dt\varphi_{Qa}\psi^a,
\eqno{(2.15)}$$

$$
\varphi_{Qa}\equiv\overrightarrow{Q}_{ab}\varphi^b,
\eqno{(2.16)}$$
for arbitrary Hermitean FODO $Q_{ab}$, and

$$
\left(\overrightarrow{\Sigma}{}^{ab}\overrightarrow{\Lambda}{}^{(0)}_{bc}
\gamma^c\right)(t)=\gamma^a(t)
\eqno{(2.17)}$$
for arbitrary function of time $\gamma^a(t)$.

We claim that $\overrightarrow{\Sigma}{}^{ab}$ is FODO. Indeed, to the lowest
order in $\varphi^\prime$ the motion equation

$$
{\delta\CS^\prime\over\delta z^a}=0
\eqno{(2.18)}$$
takes the form

$$
\overrightarrow{\Lambda}{}^{(0)}_{ab}z^b=0
\eqno{(2.19)}$$
and has the unique solution $z^a=0$. This implies that
$\hbox{det}\overrightarrow{\Lambda}{}^{(0)}_{ab}$ is a number (not a function
of the time--differentiation operators), and, hence, the matrix
$\overrightarrow{\Lambda}{}^{(0)}_{ab}$ and its inverse as well are both FODO.
Thus one can write down

$$
Y^{(n+1)a}=\mu^{(n+1)}{}^a_b(x,z)z^b
\eqno{(2.20)}$$
with $\mu^{(n+1)}{}^a_b$ being FODO.

It is easy to check the action $\CS^\prime(x,z)$, in terms of $Y^{(n+1)a}$, to
have the form

$$
\CS^\prime(x,z)=\CS_1(x)+{1\over2}\int
dt[Y^{(n+1)a}\Lambda^{(0)}_{ab}Y^{(n+1)b}+
z^a\delta_{n+2}\Lambda_{ab}(x,z)z^b]
\eqno{(2.21)}$$
with $\delta_{(n+2)}\Lambda_{ab}$ of the (n+2)-th order in
$\varphi^\prime$, being some FODO. As the assumption is obviously fulfilled to
the lowest order in $\varphi^\prime$ ($n=0$, ${\mu^{(0)}}^a_b=\delta^a_b$), one
should apply the induction method to complete the Proof.

The inverse of (2.6) has the form

$$
z^a=\tilde{\mu}^a_b(x,Y)Y^b
\eqno{(2.22)}$$
with $\tilde{\mu}^a_b$ being some FODO. Notice that, formally, the functional
determinant

$$
\hbox{Det}{\delta\varphi^{\prime A}(t^\prime)\over\delta\varphi^B(t)}
\eqno{(2.23)}$$
can be represented in the form

$$
\hbox{Det}{\delta\varphi^{\prime A}(t^\prime)\over\delta\varphi^B(t)}=
\exp[\hbox{Sp}\ln{\delta Y^a(x(t^\prime),z(t^\prime))\over\delta z^b(t)}]=
\exp[\delta(0)\int dtF(x,z)],
\eqno{(2.24)}$$
where $F(x,z)$ is a local functional of $x$, $z$ and their time--derivatives
(of a finite order to each finite order in $\varphi$).

Lemma 2. Given any solution to the quantum master equation for the theory with
classical action $\CS_1(x)$, then a solution, physically--equivalent to the
given
one, does exist to satisfy the quantum master equation for the theory with
classical action $\CS(x,z)$.

Proof. Let $\Phi^I_1=(x^i,c^\alpha,\bar{c}_\alpha,B_\alpha)$ be the complete
set of variables required to the Lagrangian BFV--quantization scheme, and let
$\Phi^*_{1I}$ be the corresponding set of antivariables. In its own turn, let
$W_1(\Phi_1,\Phi^*_1)$ be a solution to the quantum master equation,

$$
\Delta_1\exp[{\imath\over\hbar}W_1]=0,
\eqno{(2.25)}$$

$$
\Delta_1=(-1)^{\varepsilon(\Phi^I_1)}\int dt{\delta\over\delta\Phi^I_1}
{\delta\over\delta\Phi^*_{1I}}.
\eqno{(2.26)}$$
For the theory with classical action $\CS(x,z)$ we construct a solution,
physically--equivalent to the one $W_1(\Phi_1,\Phi^*_1)$, in the following way.

We start from the theory, described in terms of the variables
$\tilde{\Phi}^{\CA}=(\tilde{\Phi}^I_1,\tilde{Y}{}^a)$,
$\tilde{\Phi}{}^*_{\CA}=(\tilde{\Phi}{}^*_{1I},\tilde{Y}{}^*_a)$, with
classical
action $\tilde{\CS}(\tilde{x},\tilde{y})$ $=$
$\CS_1(\tilde{x})+{1\over2}\tilde{Y}{}^a\Lambda^{(0)}_{ab}\tilde{Y}{}^b$.

Next, we choose the solution

$$
\tilde{W}=W_1(\tilde{\Phi}_1,\tilde{\Phi}^*_1)+
{1\over2}\tilde{Y}{}^a\Lambda^{(0)}_{ab}\tilde{Y}{}^b
\eqno{(2.27)}$$
to satisfy the quantum master equation

$$
\tilde{\Delta}\exp[{\imath\over\hbar}\tilde{W}]=0, \quad \tilde{\Delta}=
\tilde{\Delta}_1+\int dt{\delta\over\delta\tilde{Y}{}^a}
{\delta\over\delta\tilde{Y}{}^*_a},
\eqno{(2.28)}$$
and boundary condition

$$
\tilde{W}\Big|_{\tilde{\Phi}{}^*=0,\hbar=0}=\tilde{\CS}.
\eqno{(2.29)}$$
By making use of the anticanonical transformation with generating function

$$
X=\int dt\left(\tilde{Y}{}^*_aY^a(x,z)+\tilde{\Phi}{}^*_{1I}\Phi^I_1\right),
\eqno{(2.30)}$$
let us pass then to the variables
$\Phi^{\CA}$ $=$ $(\Phi^I_1$, $z^a$, $\Phi^*_{1I}$, $z^*_a)$, which change
takes the following explicit form

$$
\Phi^I_1=\tilde{\Phi}{}^I_1, \quad \tilde{Y}{}^a=Y^a(x,z),
\eqno{(2.31)}$$

$$
\Phi^*_{1I}=\tilde{\Phi}{}^*_{1I}, \quad \hbox{for} \quad \Phi^*_{1I}\neq
x^*_i,
\eqno{(2.32)}$$

$$
z^*_a=\overrightarrow{Y}{}^b_a\tilde{Y}{}^*_b, \quad
\tilde{Y}{}^*_a=\overrightarrow{Z}{}^b_az^*_b,
\eqno{(2.33)}$$

$$
\tilde{x}{}^*_i=x^*_i-\overrightarrow{Y}{}^a_i\overrightarrow{Z}{}^b_az^*_b,
\eqno{(2.34)}$$
with differential operators $\overrightarrow{Y}{}^a_b$,
$\overrightarrow{Y}{}^a_i$, $\overrightarrow{Z}{}^a_b$ to be determined
from the relations

$$
\overrightarrow{Y}{}^a_b\tilde{Y}{}^*_a(t)\equiv{\delta\over\delta z^a(t)}X,
\quad
\overrightarrow{Z}{}^a_b\overrightarrow{Y}{}^b_c=\delta^a_c,
\eqno{(2.35)}$$

$$
\overrightarrow{Y}{}^a_i\tilde{Y}{}^*_a(t)\equiv{\delta\over\delta x^i(t)}X
-\tilde{x}{}^*_i(t),
\eqno{(2.36)}$$
which, in their own turn, imply all
these operators to be, perturbatively, FODO. What is more, the
change (2.31) - (2.34) and its inverse as well are both
perturbatively--local.  It has been shown in Refs. [11, 12] that given
$W(\Phi,\Phi^*)$ ($\Phi$ is a complete set of variables) to satisfy the quantum
master equation, then the action $W_X(\Phi,\Phi^*)$:

$$
W_X(\Phi,\Phi^*)\equiv W(\Phi_X,\Phi^*_X)-\imath\hbar\Delta_X,
\eqno{(2.37)}$$

$$
\Delta_X={1\over2}\ln\hbox{SDet}
{\delta(\Phi_X,\Phi^*_X)\over\delta(\Phi,\Phi^*)}
=\ln\hbox{SDet}{\delta\Phi_X\over\delta\Phi},
\eqno{(2.38)}$$
with $\Phi_X$, $\Phi^*_X$ being an anticanonical transform of $\Phi$, $\Phi^*$,
does the same.

In the case under consideration we choose (in terms of the
variables $\Phi=$$(\Phi_1,z)$)

$$
W(\Phi,\Phi^*)\equiv\tilde{W}\left(\tilde{\Phi}(\Phi,\Phi^*),\tilde{\Phi}{}^*
(\Phi,\Phi^*)\right)-\imath\hbar\hbox{Det}
{\delta\tilde{\Phi}^{\CA}(\Phi,\Phi^*)\over\delta\Phi^{\CA^\prime}}.
\eqno{(2.39)}$$
to be a "particular" solution to the quantum master equation

$$
\Delta\exp[{\imath\over\hbar}W]=0, \quad \Delta=\Delta_1+
\int dt{\delta\over\delta z^a}{\delta\over\delta z^*_a}.
\eqno{(2.40)}$$
At the same time the solution chosen obviously satisfies the boundary condition

$$
W\Big|_{\Phi^*=0,\hbar=0}=\CS(x,z).
\eqno{(2.41)}$$
Let us check the actions $W_1(\Phi_1,\Phi^*_1)$ and $W(\Phi,\Phi^*)$ to be
physically--equivalent. In the theory, parametrized by the mentioned variables
$\Phi$, $\Phi^*$ let us choose the gauge fermion $\Psi$ to depend on the
variables $\Phi_1$ only:

$$
\Psi=\Psi(\Phi_1).
\eqno{(2.42)}$$
Then we have for the statsum $Z$:

$$
Z=\int D\Phi\exp[{\imath\over\hbar}W\Big|_{\Phi^*_1=
{\partial\Psi\over\partial\Phi_1},z^*=0}]=
\int D\Phi\exp[{\imath\over\hbar}W_1\Big|_{\Phi^*_1=
{\partial\Psi\over\partial\Phi_1}}]=Z_1.
\eqno{(2.43)}$$
It can be also easily checked that, to the first order in $z^*$, the expression
(2.30), taken at $\hbar=0$, coincides with the action given in Ref. [17].

The inverse of the Lemma 2 is valid too.

Lemma 3. Given any solution $W(\Phi,\Phi^*)$ to the quantum master equation
(2.40), then a physically--equivalent solution $W_1(\Phi_1,\Phi^*_1)$ does
exist
to satisfy the quantum master equation (2.35), (2.36).

Proof. By choosing the gauge fermion $\Psi$ to depend on $\Phi_1$ only, we have

$$
\exp[{\imath\over\hbar}W_1(\Phi_1,\Phi^*_1)]\equiv\int Dz
\exp[{\imath\over\hbar}W\Big|_{z^*=0}]=\int
DY\exp[{\imath\over\hbar}\bar{W}], \eqno{(2.44)}$$

$$
\bar{W}=W\Big|_{z^*=0}+\imath\hbar\hbox{Det}{\delta Y^a\over\delta z^b}=
\CS_1(x)+{1\over2}Y^a\Lambda^{(0)}_{ab}Y^b+M,
\eqno{(2.45)}$$
where $M=0$ at $\Phi^*_1=0$, $\hbar=0$.

Obviously, $W_1$ is, formally, a local functional of $\Phi_1$, $\Phi^*_1$ , and

$$
W_1\Big|_{\Phi^*_1=0,\hbar=0}=\CS_1(x).
\eqno{(2.46)}$$
Besides, by making use of the method of Ref. [13], one can check $W_1$ thus
defined to satisfy the quantum master equation. We conclude the Section with
the following claim: if the actions $\CS$ and $\CS_1$ are
classically--equivalent
in the sense of the relations (2.2), (2.3), then the corresponding
BV--quantized
theories are also physically--equivalent in the sense that the existence of a
solution to the quantum master equation in the first theory implies the
existence of a physically--equivalent solution to the quantum master equation
in the second theory, and  vice versa as well.

\section{Proper Hamiltonian action classically--equivalent to Lagrangian one}

Let us begin here with the Lagrangian action

$$
\CS_L\equiv\int dtL(q,\dot{q}).
\eqno{(3.1)}$$
Without a loss of generality one can suppose the Lagrangian $L$ to depend only
on coordinates and velocities.

Next, let us introduce [18] the action $\CS_v$:

$$
\CS_v=\int dt(L(q,v)+p(\dot{q}-v)).
\eqno{(3.2)}$$
The action $\CS_v$ is classically--equivalent to the  one $\CS_L$,
as one obtains $\CS_L$ by substituting $v$ $\rightarrow$
$v(q,\dot{q})=\dot{q}$, $p$ $\rightarrow$ $p(q,\dot{q})=\partial
L/\partial\dot{q}$ into $\CS_v$, which substitution, in its own
turn, is determined by the equations

$$
{\delta\CS_v\over\delta p}={\delta\CS_v\over\delta v}=0.
\eqno{(3.3)}$$
Let the velocities $v$, as well as the momenta $p$, are split into the
corresponding subsets

$$
v=(V,\lambda), \quad p=(\Pi,\pi)
\eqno{(3.4)}$$
in such a way that the submatrix

$$
{\partial^2L\over\partial V\partial V}
\eqno{(3.5)}$$
is the maximal--size square block of the Hessian

$$
{\partial^2L\over\partial v\partial v}
\eqno{(3.6)}$$
whereas the momenta $\Pi$ are determined from the equations

$$
{\delta\CS_v\over\delta V}={\partial L\over\partial V}-\Pi=0.
\eqno{(3.7)}$$
These equations are solvable with respect to $V$'s:

$$
(3.7) \quad \Rightarrow \quad V=\bar{V}(\Pi,\lambda,q).
\eqno{(3.8)}$$
By substituting these $V$'s into the action $\CS_v$ one obtains
the action $\CS_H$ classically--equivalent to the action $\CS_v$
(and hence to $\CS_L$):

$$
\CS_H=\CS_v\Big|_{V\to\bar{V}}=\int dt(p\dot q-H(\Pi,q)-\lambda\Phi^{(1)}),
\eqno{(3.9)}$$
with $\Phi^{(1)}=\pi-f(\Pi,q)$ being primary constraints. It has been shown in
Ref.[18] that there exists a point--like change of variables ($p$, $q$,
$\lambda$) $\to$ ($p^\prime$, $q^\prime$, $\lambda^\prime$), which is a
canonical transformation as applied to the variables $p$, $q$, such that the
action $\CS_H$ takes the form (we omit primes)

$$
 \CS_H \quad \to \quad \CS^\prime_H=\int dt\left(p\dot q-H_{ph}(\omega)-
A(\omega,\SP^{(2)},Q)\SP^{(2)}-\Delta H-\lambda_{\SP}\SP^{(1)}-
\lambda_\theta\theta^{(1)}\right)
 \eqno{(3.10)}$$
where $\eta^\prime=(p^\prime,q^\prime)=(\omega;\SP,Q;\theta)$, $\theta$ is a
set of canonical pairs describing second--class constraints, $\SP$ is a set
of momenta describing first--class constraints, $Q$ is a set of coordinates
canonically--conjugated to $\SP$, $\omega$ is a set of canonical pairs
describing physical degrees of freedom, $\theta^{(1)}$ is a set of primary
second--class constraints, $\SP^{(1)}$ is a set of primary first-class
constraints, $\lambda_{\SP}$ and $\lambda_\theta$ are Lagrange multipliers,
$\theta^{(2)}$ represent all secondary second--class constraints (the ones of
the second, third, etc. steps of the Dirac's procedure), $\SP^{(2)}$
represent all secondary first-class constraints. The function $\Delta H$ has
the structure

$$
\Delta H=O\left(\SP^{(2)}\theta^{(2)},(\theta^{(2)})^2\right)
\eqno{(3.11)}$$
and can be canonically transformed to become

$$
\Delta H={1\over2}\theta^{(2)}b\theta^{(2)}+\Delta_3H
\eqno{(3.12)}$$
where $\Delta_3H$ is of the third order in $\eta^\prime$. Let us denote
$\bar{\theta}$ $=$ $(\theta,\lambda_\theta)$. The equations of motion coming
from $\CS^\prime_H$ by varying with respect to $\bar{\theta}$ have the
structure

$$
B\bar{\theta}={\partial\Delta_3H\over\partial\bar{\theta}}
\eqno{(3.13)}$$
where $B$ is a matrix differential (in time) operator entering a quadratic
theory. As $\bar{\theta}=0$ is the only solution in a quadratic theory,
$\det B$ is a number, and $B^{-1}$ is FODO. The eq. (3.13) determines
$\bar{\theta}$ to be some functions of $\omega,\SP,Q$ and their
time-derivatives
of a finite order, to each finite order in $\SP$. In other words,
$\bar{\theta}$
are perturbatively--local functionals (i.e. functions) of $\omega,\SP,Q$ and
their time--derivatives, such that $\bar{\theta}\Big|_{\SP=0}=0$.

Let us substitute the expressions obtained for $\bar{\theta}$ into
$\CS^\prime_H$. As a result, one gets the action $\CS^{\prime\prime}_H$
classically--equivalent to $\CS^\prime_H$ (and, hence, to $\CS_L$), which has
the structure

$$
\CS^{\prime\prime}_H=\int dt\left(\omega_p\dot\omega_q+\SP\dot Q-H_{ph}(\omega)
-A(\omega,\SP^{(2)},Q)\SP^{(2)}-\lambda_{\SP}\SP^{(1)}+C\right),
\eqno{(3.14)}$$
where $C$ is a local functional of $\omega,\SP,Q$ and their time--derivatives,
which is at least quadratic in $\SP$.

The action $\CS^{\prime\prime}$ does not depend on the variables
$\theta$, $\lambda_\theta$ which correspond to the second--class constraints.
This action, however, is not of the Hamiltonian form, since the function $C$
may depend on the time--differentiated phase variables. We are going to show
that there exists a perturbatively--local change of variables that eliminates
the function $C$.

For the sake of simplicity we restrict ourselves by a nonprincipal extra
assumption that the same number of first-class constraints appears at each step
 of the Dirac procedure, which number equals to the one of primary first--class
 constraints. Thus we suppose that

$$
\SP=(\SP_{(1)},\SP_{(2)},\ldots,\SP_{(L)}), \quad \SP_{(1)})\equiv\SP^{(1)}.
\eqno{(3.15)}$$
One can show the contribution $A\SP^{(2)}$ to take the form

$$
\begin{array}{c}\displaystyle
A(\omega,\SP^{(2)},Q)\SP^{(2)}=A(\omega,\SP^{(2)},0)\SP^{(2)}+\sum_{i=2}^L
\left(Q_{(i-1)}+\sum_{k=1}^LQ_{(k)}a^{(k)(i)}(\omega)\right)\SP_{(i)}+
\\ [9pt] \displaystyle
+O(Q(\SP^{(2)})^2,Q^2\SP^{(2)}), \quad a^{(k)(i)}(\omega)=O(\omega).
\end{array} \eqno{(3.16)}$$
as a result of a (linear) canonical transformation.

Now let us suppose the change

$$
\bar{Q}_{(i)} \to \bar{Q}{}^\prime_{(i)}, \quad Q_{(L)} \to
Q^\prime_{(L)}=Q_{(L)}, \quad \SP \to \SP^\prime=\SP, \quad \omega
\to \omega^\prime=\omega,
\eqno{(3.17)}$$
where

$$
\bar{Q}_{(1)}=\lambda_{\SP}, \quad \bar{Q}_{(i)}=Q_{(i-1)}, \quad
i=2,\ldots,L,
\eqno{(3.18)}$$
to make the action $\CS^{\prime\prime}_H$ take the form (3.14) with the
function
$C$ being of the minimal order $n\ge2$ in $\SP$. By integrating over time by
part one can free one of the factors $\SP$ from the time--differentiation,
so that the $C$--contribution takes the form

$$
\int dt\left(\SP_{(i)}T^{(i)}+O(\SP^{n+1})\right), \quad
T^{(i)}=O(\SP^{n-1}).
\eqno{(3.19)}$$
Next, let us make the change

$$
\bar{Q}_{(i)} \quad \to \bar{Q}^\prime_{(i)}=\bar{Q}_{(i)}+\Delta_{(i)},
\eqno{(3.20)}$$
where $\Delta_{(i)}$ are determined by the equations

$$
A^{(i)(j)}_1\Delta_{(j)}+A^{(i)(j)}_2(\omega)\Delta_{(j)}=T^{(i)},
\eqno{(3.21)}$$

$$
A^{(i)(j)}_1=\delta_{ij}-\delta_{i,j-1}{d\over dt},
\eqno{(3.22)}$$

$$
A^{(1)(j)}_2(\omega)=A^{(i)(1)}_2(\omega)=0; \quad
A^{(i)(j)}_2(\omega)= a^{(i)(j-1)}(\omega), \quad i,j\ge2.
\eqno{(3.23)}$$
As the operator $A^{-1}_1$ is FODO, the operator $(A_1+A_2)^{-1}$ does
perturbatively the same, and $\Delta_{(j)}$ are local functionals of $\omega$,
$\SP$, $Q$ and their time--derivatives of finite orders.

As a result of such a change the action $\CS^{\prime\prime}_H$ takes the form
(3.14) with the function $C$ being of the order $\sim$ $\SP^{n+1}$.

Finally we conclude the initial action $\CS_L$ to be dynamically--equivalent
to the one $\CS_{1H}$ (up to an invertible local in time change of variables):

$$
\CS_{1H}=\CS_H\Big|_{\theta=\lambda_\theta=0}=\int dt(\omega_p\dot\omega_q+
\SP\dot Q-H_{ph}(\omega)-A\SP^{(2)}-\lambda_{\SP}\SP^{(1)}).
\eqno{(3.24)}$$
The action $\CS_{1H}$ is of the Hamiltonian form, and generates first-class
constraints only.

\section{Solving quantum master equation}

In this Section we construct the solution $W_1$ to the quantum master equation
for the theory with classical action $\CS_{1H}$ (3.24), which solution is
physically--equivalent to the Dirac quantization of the theory (3.24). As the
Dirac method applied to the theory (3.24) yields the answer

$$
Z=\int D\omega\exp[{\imath\over\hbar}\int dt(\omega_p\dot\omega_q-
H_{ph}(\omega))
\eqno{(4.1)}$$
which coincides with the Dirac quantization of the initial theory (3.1), we
thereby guarantee that the quantum theory with the action $W_1$ is
physically--equivalent to the theory quantized by applying the Dirac method
and,
hence, to the BV--quantized theory.

To construct $W_1$ we make use of the scheme suggested in Ref. [14]. Let us
split the complete set of phase variables $\Gamma$ of the BFV--quantization
applied to the theory (3.24),

$$
\Gamma=\left(\omega; \quad \SP_{(i)},Q_{(i)}; \quad
C_{(i)},\BCP_{(i)}; \quad \bar{C}_{(i)},\CP_{(i)};
\quad \lambda_{(i)},\pi_{(i)}\right), \quad i=1,\ldots,L,
\eqno{(4.2)}$$
into the two groups $x$ and $y$, where

$$
x=(\omega, \quad \SP, \quad Q, \quad \lambda\equiv\lambda_{(1)}
\equiv\lambda_{\SP}, \quad
C\equiv C_{(L)},\quad\bar{C}\equiv\bar{C}_{(L)}, \quad B\equiv\pi_{(L)}),
\eqno{(4.3)}$$
and the rest of the variables is included into the group $y$.

Let us define the unitarizing Hamiltonian via the formula

$$
H=\CH_{min}(\omega,\SP,Q,C_{(i)},\BCP_{(i)})+\{\Psi,\Omega\},
\eqno{(4.4)}$$
where

$$
\Omega=\Omega_{min}(\omega,\SP,Q,C_{(i)},\BCP_{(i)})+\pi_{(i)}\CP_{(i)},
\eqno{(4.5)}$$
the Fermionic $\Omega_{min}$ and Bosonic $\CH_{min}$ generating functions of
the
constraint algebra satisfy the generating equations

$$
\{\Omega_{min},\Omega_{min}\}=0, \quad \Omega_{min}=\SP_{(i)}C_{(i)}+
O(\BCP C^2),
\eqno{(4.6)}$$

$$
\{\CH_{min},\Omega_{min}\}=0, \quad \CH_{min}=H_{ph}(\omega)+\sum_{i=2}^L
A(\omega,\SP^{(2)},Q)^{(i)}\SP_{(i)}+\BCP_{(i)}V^{(i)(j)}C_{(j)}+O(C^2\BCP^2),
\eqno{(4.7)}$$
and

$$
V^{(i)(j)}={\partial A^{(i)}\over\partial Q_{(j)}}
\eqno{(4.8)}$$
Let us choose the gauge Fermion $\Psi$ to have the form

$$
\Psi=-\Psi_{s}(x)+\Psi_1(y)
\eqno{(4.9)}$$
where

$$
\Psi_1(y)=\BCP_{(1)}\lambda+b^2\sum_{i=1}^{L-1}\bar{C}_{(i)}(\pi_{(i)}+
\lambda_{(i+1)}).
\eqno{(4.10)}$$
Then let us define the action $\tilde{W}_b(x,x^*)$ to be

$$
\exp[{\imath\over\hbar}\tilde{W}_b(x,x^*)]=\int Dy\exp[{\imath\over\hbar}
\int dt(\Gamma_p\dot\Gamma_q-\tilde{H})]
\eqno{(4.11)}$$
where

$$
\tilde{H}=H-x^*\{x,\Omega\}=\CH_{min}+\{\Psi_1,\Omega\}-(x^*+
{\partial\Psi_s\over\partial x})\{x,\Omega\}.
\eqno{(4.12)}$$
The action $\tilde{W}_b$ has the two important properties:

i) $\tilde{W}_b$ satisfies [13] the quantum master equation

$$
\Delta_x\exp[{\imath\over\hbar}\tilde{W}_b]=0, \quad
\Delta_x=(-1)^{\varepsilon(x)}\int dt{\delta\over\delta x}
{\delta\over\delta x^*};
\eqno{(4.13)}$$

ii) $x^*$ and $\Psi_s$ enter $\tilde{W}_b$ only through the combination
$\bar{x}^*\equiv x^*+\partial\Psi_s/\partial x$:

$$
\tilde{W}_b(x,x^*)=W_b(x,\bar{x}{}^*).
\eqno{(4.14)}$$
It is thus sufficient for our purposes to analyze the action $W_b(x,x^*)$:

$$
\exp[{\imath\over\hbar}W_b(x,x^*)]=\int Dy\exp[{\imath\over\hbar}
\int dt(\Gamma_p\dot\Gamma_q-\CH_{min}-\{\Psi_1,\Omega\}+x^*\{x,\Omega\})].
\eqno{(4.15)}$$
Let us make the unimodular change of the integration variables

$$
\lambda_{(i)} \quad \to \quad {1\over b}\lambda_{(i)},\quad
\CP_{(i)} \quad \to \quad {1\over b}\CP_{(i)}, \quad i\ge2,
\eqno{(4.16)}$$

$$
\pi_{(i)} \quad \to \quad {1\over b}\pi_{(i)}, \quad
\bar{C}_{(i)} \quad \to \quad {1\over b}\bar{C}_{(i)}, \quad i\le
L-1, \eqno{(4.17)}$$ and then consider the limit $b\to\infty$.
We have

$$
\begin{array}{c} \displaystyle
\exp[{\imath\over\hbar}W_\infty]= \\[9pt] \displaystyle
=\exp[{\imath\over\hbar}(\CS_{1H}+\int dtC^*B)]
\int\prod_{i\ge2}D\BCP_{(i)}\prod_{k\le L-1}DC_{(k)}
\exp[{\imath\over\hbar}\int dt\Bigl(\BCP_{(L)}\dot C+
\sum_{i=2}^{L-1}\BCP_{(i)}\dot C_{(i)}- \\[9pt] \displaystyle
-\sum_{i=2}^{L}\sum_{j=1}^{L-1}\BCP_{(i)}V^{(i)(j)}C_{(j)}+
O(\lambda C_{(i)}\BCP_{(j)},C_{(i)}C_{(j)}\BCP_{(k)}\BCP_{(l)})+
x^*_{min}\{x_{min},\Omega_{min}\}_0\Bigr)],
\end{array} \eqno{(4.18)}$$
where

$$
x_{min}=(\omega;\SP,Q;\lambda;C),
\eqno{(4.19)}$$

$$
\{x_{min},\Omega_{min}\}_0\equiv\{x_{min},\Omega_{min}\}\Big|_{\BCP_{(1)}=0}
\eqno{(4.20)}$$
It follows from (4.18) that $W_{\infty}$ has the structure

$$
W_{\infty}(x,x^*)=W_{min}(x_{min},x^*_{min})+\int dtC^*B,
\eqno{(4.21)}$$
where $W_{min}$, obviously, satisfies the quantum master equation in the
variables $x_{min}$, $x^*_{min}$, together with the boundary condition

$$
W_{min}\Big|_{x^*_{min}=0,\hbar=0}=\CS_{1H}.
\eqno{(4.22)}$$
Thus, given the classical action $\CS_{1H}$, the action $\tilde{W}_{\infty}$ is
constructed by the standard rules of the Lagrangian BV--quantization, so that
the theory thus defined is physically--equivalent to the theory (3.1) quantized
by applying the Dirac method. It remains to check the action $W_{min}$ to be
local.

Let us represent a quadratic in $\BCP$, $C$ contribution to the action
in the exponent of the integrand of eq. (4.18) in the form (see also (3.16)):

$$
\int dt(\sum_{i=2}^{L-1}\BCP_{(i)}\dot C_{(i)}-
\sum_{i=2}^{L}\BCP_{(i)}C_{(i-1)})\equiv
\int dtC_{(i)}\sigma^{(i)(j)}\BCP_{(j)}, \quad i\ge2, j\le L-1.
\eqno{(4.23)}$$
Notice that, to the linear order in the variables, the equations of motion,
which follow from the action $\CS_{1H}$ by varying with respect to
$\bar{Q}_{(i)}$, have the form

$$
\sum_{j=1}^LA^{(i)(j)}\SP_{(j)}=0, \quad A^{(i)(j)}=
\left(\begin{array}{cc}1&0\\
{}*&\sigma\end{array}\right).
\eqno{(4.24)}$$
As $\SP=0$ is the only solution to the equations (4.25), the
time--differentiation operators do not enter $\det A=\det\sigma$, so that
$\sigma^{-1}$ is FODO (this follows also from the explicit form of the operator
$\sigma$) and, hence, the propagators of the fields $C_{(i)}$, $i\le L-1$, and
$\SP_{(j)}$, $j\ge2$ are operators local in time.

So we have finally shown that there exists a local solution to the quantum
master equation for the theory with classical action $\CS_{1H}$, which solution
is physically--equivalent to the quantization based on the Dirac
formalism. As the initial action $\CS_L$ is classically--equivalent to the one
$\CS_{1H}$, we conclude that a solution to the quantum master equation does
exist for the initial theory too, which solution is physically--equivalent to
the Dirac scheme. In that sense the Hamiltonian and Lagrangian quantizations
are
equivalent. Besides, it follows from the above consideration that, in general,
the quantum master equation has, perturbatively, a local solution to satisfy a
given boundary condition as well.

{\bf Acknowledgement.} This work was supported in part by the
Russian Foundation of Fundamental Investigations, Grant \# 93--02--15541.

\end{document}